\font\espa=cmss10
\font\chi=cmmi7

\def\ah{\hbox{\bf A}} 
\def\ce{\hbox{\espa\char67}}
\def\dcal{\hbox{\cal\char68}}
\def\dh{\hbox{\bf D}}
\def\dhk{\dh\lower .5ex\hbox{\kern.0em{\chi k}}}
\def\dhx{\dh\lower .5ex\hbox{\kern.0em{\chi x}}}
\def\dhy{\dh\lower .5ex\hbox{\kern.0em{\chi y}}}
\def\dhtet{\dh\lower .5ex\hbox{\kern.0em{\chi\char18}}}
\def\dhfi{\dh\lower .5ex\hbox{\kern.0em{\chi\char39}}}
\def\dhr{\dh\lower .5ex\hbox{\kern.0em{\chi r}}}
\def\ere{\hbox{\espa\char82}} 
\def\eh{\hbox{\bf E}} 
\def\lh{\hbox{\bf L}}
\def\fh{\hbox{\bf f}}

\def\gh{\hbox{\bf g}}
\def\ph{\hbox{\bf P}}
 
\def\sh{\hbox{\bf S}} 

\def\taug{\hbox{$\tau$}}
\def\teth{\hbox{$\bf\theta$}}
\def\xh{\hbox{\bf X}}
\def\xhk{\xh\raise 1.1ex\hbox{\kern.15em{\chi k}}}
\def\yh{\hbox{\bf Y}}

\parindent 1truecm
\null\vskip .2truecm
\centerline{\bf  A finite--dimensional representation of the quantum 
angular momentum operator}
\bigskip
\centerline{(Short title: Angular momentum in a finite linear space)}
\vskip 1truecm
\centerline{Rafael G. Campos}
\vskip .15truecm
\centerline{Escuela de Ciencias F\'{\i}sico--Matem\'aticas,
Universidad Michoacana}
\vskip .15truecm
\centerline{58060 Morelia, Michoac\'an, M\'exico}
\vskip .15truecm
\centerline{\tt e-mail: rcampos@zeus.ccu.umich.mx}
\vskip .5truecm
\centerline{L. O. Pimentel}
\vskip .15truecm
\centerline{Departamento de F\'{\i}sica, Universidad Aut\'onoma
Metropolitana--Iztapalapa}
\vskip .15truecm
\centerline{Apdo. Postal 55-534, M\'exico DF, 09340 M\'exico}
\vskip .15truecm
\centerline{\tt e-mail: lopr@xanum.uam.mx}
\vskip .9truecm
\leftline{\bf  Abstract}
\vskip .4truecm
A useful finite--dimensional matrix representation of the derivative of 
periodic functions is obtained by using some elementary facts of 
trigonometric interpolation. This $N\times N$ matrix becomes a 
projection of the angular derivative 
into polynomial subspaces of finite dimension and it 
can be interpreted as a generator of discrete rotations  
associated to the $z$--component of the projection of the angular momentum 
operator in such subspaces, inheriting thus some properties of the 
continuum operator. The group associated to these discrete rotations 
is the cyclic group of order $N$.\par
Since the square of the quantum angular momentum $L^2$ is associated to 
a partial differential boundary value problem in the angular variables 
$\theta$ and $\varphi$ whose solution is given in terms of the spherical 
harmonics, we can project such a differential equation to obtain an 
eigenvalue matrix problem of finite dimension by extending to several 
variables a projection technique for solving numerically two point boundary 
value problems and using the matrix representation of the angular derivative 
found before. The eigenvalues of the matrix representing $L^2$ are found to 
have the exact form $n(n+1)$, counting the degeneracy, and the eigenvectors 
are found to coincide exactly with the corresponding spherical harmonics 
evaluated at a certain set of points.\par
\vskip 1truecm
\hrule
\medskip
\leftline{PACS: 03.65.Ca, 03.65.Lx, 02.60.Jh, 02.60.Lj}
\noindent
\null\bigskip
\leftline{\bf 1. Introduction}\par
\parindent 1truecm
A Galerkin--collocation--type method based in a $N$--dimensional 
matrix representation of ${{d}\over{dx}}$ obtained through Lagrange's 
interpolation has been used to solve one--dimensional boundary value 
problems (see Refs. [1]--[4]). This technique consists basically in 
the substitution of the variable $x$ and the derivative ${{d}\over{dx}}$ 
by $N\times N$ matrices $X$ and $D$ in a certain form of the given 
differential equation. The diagonal matrix $X$ has the $N$ different 
values $x_j$, $j=1,2,\cdots,N$ along the main diagonal, and the matrix 
$D$ is given by
$$D=P\tilde{D}P^{-1},\leqno(1)$$
where
$$\tilde{D}_{ij}=\cases{\displaystyle\mathop{{\sum}'}\limits_{l=1}^N
{1\over(x_i-x_l)},&{$i=j$},\cr\noalign{\vskip .5truecm}
\displaystyle {1\over(x_i-x_j)}, &{$i\not=j$},\cr}\hskip 1.7truecm
\displaystyle P_{ij}=p'(x_i)\delta_{ij},\qquad i,j=1,2,\cdots,N,\leqno(2)$$
The symbol $\sum'$ appearing in (2) means the sum over $l\ne i$ and
the prime on $p$ means differentiation of the polynomial
$$p(x)=\prod_{k=1}^N(x-x_k).$$
The $N$ nodes $x_j$ can be chosen by imposing
a condition where the coefficients of the differential equation
and the boundary conditions play the main part. This condition is
$$\mathop{{\sum}'}\limits_{k=1}^N{1\over(x_j-x_k)}
=-{{\gamma'(x_j)}\over{\gamma(x_j)}},\qquad j=1,2,\cdots,N,\leqno(3)$$
\noindent
where $\gamma(x)$ is a function defined by the boundary conditions and
the differential equation (see Refs. [4]--[5]).\par
More precisely, 
the projection scheme for the $k$--th derivative of a real function $f$
evaluated at different (but otherwise arbitrary) points 
$x_i, i=1,2,\ldots,N$ is given by
$$f^{(k)}(x_j)=\sum^N_{l=1}{D^k}_{jl}f(x_l)+E_j,
\qquad j=1,2,\cdots,N,\leqno(4)$$
where $E_j$ is the $j$--th component 
of the residual vector depending on $f$, $N$ and $k$. Let us denote 
by $\pi_{N-1}$ the space of polynomials of degree at most $N-1$. Thus, 
if $f\in\pi_{N-1}$,
it is found that (4) is exact at the nodes, {\it i.e.}, $E_j=0$, and 
therefore, any differential problem closed in $\pi_{N-1}$
can be solved also in ${\ere}^N$ as a matrix problem yielding the same 
solution. An estimate of the error $E_j$ for other kind of functions is 
given in [4].\par
The formal application of this elementary technique to some multivariate 
problems is straightforward, as shown in the following section (see also
[6]).\par
\bigskip\goodbreak
\noindent
\leftline{\bf 2. Multivariate case}\par
To illustrate how this method should be applied to some boundary value 
problems in several variables we will take first two real 
variables, $x$ and $y$.\par
Let $\tilde{\Pi}=\pi_{M-1}\otimes\pi_{N-1}$ be
the space of bivariate tensor--product polynomials of degree $N-1$ in
the variable $x$ and $M-1$ in $y$. Thus, if $f\in\tilde{\Pi}$, $f(x,y)$ 
can be written as
$$f(x,y)=\sum_{j=0}^{N-1}\sum_{k=0}^{M-1}a_{jk}x^jy^k.\leqno(5)$$
Now let us take two sets of different but otherwise arbitrary points
$\{x_1,x_2,\cdots,x_{N_1}\}$ and $\{y_1,y_2,\cdots,y_{N_2}\}$,
on the $x$ and $y$ axes respectively, and let $D_x$ and $D_x$ be the 
$N\times N$ and $M\times M$ matrix representations of ${{d}\over{dx}}$ 
and ${{d}\over{dy}}$ constructed according to (1) with such sets.\par
By deriving partially (5) $n$ times
with respect to $x$ and $m$ times with respect to $y$,  
and evaluating the result at the cartesian nodes
$(x_j,y_k)$
we obtain the vector ${\fh}^{\kern.15em (n,m)}$ of dimension 
$\tilde{N}=NM$ whose entries are given by
$$f^{(n,m)}_r={{\partial^{n+m}f(x,y)}\over{\partial x^n
\partial y^m}}\Big\vert_{(x_j,y_k)}$$
where the indexes $r$, $j$ and $k$ are related through
$$r=j+(k-1)N,\quad j=1,2,\cdots,N,\quad k=1,2,\cdots,M,\leqno(6)$$
in such a form that $r=1,2,\cdots,\tilde{N}$. 
It is known that a bivariate interpolation on the grid 
$(x_j,y_k)$ is uniquely possible in $\tilde{\Pi}$ (see [7]) 
so that $f$ can be taken as a bivariate and sufficiently 
differentiable real function other than
a polynomial, and (5) as its corresponding (tensor) Taylor polynomial.
By choosing $j$ (the $x$--index) to run faster than
$k$ and using (4) according to the case, we can write down a matrix 
formula in ${\ere}^{\tilde{N}}$ for ${\fh}^{\kern.15em (n,m)}$ in terms of 
${\fh}={\fh}^{\kern.15em (0,0)}$, generalizing (4):
$${\fh}^{\kern.15em (n,m)}=({D_y}^m\otimes {D_x}^n){\fh}+{\eh}.\leqno(7)$$
Here, the Kronecker product of matrices $A=(a_{jk})$ and $B=(b_{jk})$ of
sizes $N\times N$ and $M\times M$, respectively, is defined by
$$A\otimes B=\pmatrix{a_{11}B&a_{12}B&\cdots&a_{1,N}B\cr
                      a_{21}B&a_{22}B&\cdots&a_{2,N}B\cr
                      \vdots&\vdots&\ddots&\vdots\cr
                      a_{N,1}B&a_{N,2}B&\cdots&a_{N,N}B\cr}.$$
The components of the residual vector {\eh} are zero if $f\in\tilde{\Pi}$ 
and an expression
for them is given in [8] for sufficiently differentiable functions.\par
Let $1_x$ and $1_y$ denote the identity matrices of dimension
$N$ and $M$ respectively, and
define the $\tilde{N}\times\tilde{N}$ matrices ${\dhx}$ and ${\dhy}$ by
$${\dhx}=1_y\otimes D_x, \qquad {\dhy}=D_y\otimes 1_x.\leqno(8)$$
Due to the properties of the Kronecker product, these matrices commute:
$${\dhx}{\dhy}={\dhy}{\dhx}=D_y\otimes D_x.$$
More generally,
$${\dhx}^n{\dhy}^m=(1_y\otimes D_x^n)(D_y^m\otimes 1_x)=
D_y^m\otimes D_x^n,$$
and therefore, (7) takes the form
$${\fh}^{\kern.15em (n,m)}={\dhx}^n{\dhy}^m{\fh}+{\eh},\leqno(9)$$
indicating that the partial derivatives
$\partial/\partial x$ and $\partial/\partial y$ 
take in $\tilde{\Pi}$ the tensor--product forms given in (8).\par
By using the properties of the tensor product and defining
${\ph}=P_y\otimes P_x$ and
$$\tilde{\dhx}=1_y\otimes \tilde{D_x}, \qquad 
\tilde{\dhy}=\tilde{D_y}\otimes 1_x,$$
where $P_x$, $P_y$, $\tilde{D_x}$ and $\tilde{D_y}$ have the structure 
given in (2),
it is possible to give the following alternate form of (9):
$${\fh}^{\kern.15em (n,m)}={\ph}{\tilde{\dhx}}^n
{\tilde{\dhy}}^m{\ph}^{-1}{\fh}+{\eh}.$$\par
On the other hand, the projection of the coefficient functions of the 
differential operator can be written as diagonal matrices since the 
the product of a function $a(x,y)$ by the partial derivatives of the 
unknown function evaluated at the nodes $(x_j,y_k)$ is  
$$a_rf^{(n,m)}_r=a(x_j,y_k)f^{(n,m)}(x_j,y_k)$$
($a(x_j,y_k)$ should be well defined) 
and the indexes can be ordered according to (6), producing that in
this scheme, the coefficient functions can be represented by 
$\tilde{N}\times\tilde{N}$ diagonal
matrices whose non-zero elements are given by $a_r=a(x_j,y_k)$,
where $r$, $j$ and $k$ are related by (6). Let us denote this generic
coefficient matrix by ${\ah}$. Thus, the part of the differential
operator consisting in the product $a(x,y)f^{(m,n)}(x,y)$
takes the matrix form 
$${\ah}{\dhx}^m{\dhy}^n{\fh}.$$
If $a(x,y)$ accept a Taylor expansion (this condition is too restrictive 
and it can be relaxed, but it is adequate for our illustrative purposes),
we have that {\ah} can be defined by the same function $a(x,y)$ through 
$${\ah}=a({\xh},{\yh}),$$ 
where {\xh} and {\yh} are the matrices given by
$${\xh}=1_y\otimes X,\qquad {\yh}=Y\otimes 1_x,$$
where the $N\times N$ diagonal matrix $X$ has the set of points $x_j$, 
$j=1,2\cdots,N$ along the main diagonal whereas the $M$ points $y_j$ lie 
along the main diagonal of the $M\times M$ matrix $Y$. {\xh} and {\yh} 
represent the variables $x$ and $y$, respectively.\par 
The generalization to the case of $q$ variables $x_1,x_2,\cdots,x_q$
is straightforward. We will choose $N_j$ points on the $x_j$--axis,
and the projection space as the tensor product of the subspaces of
polynomials of degree at most $N_j-1$ in $x_j$, {\it i.e.},
$$\tilde{\Pi}=\mathop{\otimes}\limits_{j=1}^q\pi_{N_j-1}.$$ 
The nodes will be ordered in such a way that a function $f(x_1,\cdots,x_q)$
evaluated at the nodes yields the vector {\fh} whose entries are
$$f_r=f(x^1_{j_1},x^2_{j_2},\cdots,x^q_{j_q}), 
\quad j_k=1,2,\cdots,N_k,\quad k=1,2,\cdots,q,$$
and the indexes $r$ and $j_k$ are related through
$$r=j_1+(j_2-1)N_1+(j_3-1)N_1N_2\cdots+(j_q-1)\tilde{N}/N_q,
\quad j_k=1,2,\cdots,N_k,\quad k=1,2,\cdots,q,$$
and $\tilde{N}=\prod_{k=1}^q N_k$. The index than runs faster is $j_1$,
then $j_2$ and so on, yielding that $r=1,2,\cdots,\tilde{N}$.\par
The $\tilde{N}\times\tilde{N}$ matrix representation of 
$\partial/\partial x^k$, is now
$${\dhk}=1_q\otimes\cdots\otimes 1_{k+1}\otimes 
D_k\otimes 1_{k-1}\cdots\otimes 1_1,$$
where $1_j$ is the $N_j\times N_j$ identity matrix and 
$D_k$ is a matrix of dimension $N_k$ having the structure given by (1).
Similarly, the representation of the variable $x^k$ is
$${\xhk}=1_q\otimes\cdots\otimes 1_{k+1}\otimes 
X^k\otimes 1_{k-1}\cdots\otimes 1_1.$$\par
In the next section we adapt this technique to other kind of subspaces
and it will be applied to an important physical problem in Section 4.\par
\bigskip\goodbreak
\noindent
\leftline{\bf 3. Discrete rotations}\par
In this section we show that a matrix representation of the angular 
derivative (the derivative of univariate periodic functions), can be 
related to a generator of discrete rotations (associated to a subgroup 
of the rotation group) and becomes a finite--dimensional matrix 
representation of the $z$--component of the angular momentum.\par
The notation and language used in this section is that of Quantum Physics, 
what can be seen as a digression in this numerical look--like 
paper, but we think that the reasons are obvious.\par
Let us begin by considering a complete set of quantum states in a
space of finite dimension, $\vert\varphi_j\rangle$, to be determined
later. Here, $\varphi_j$ indicates the $j$--th eigenvalue of the 
operator associated to the spatial observable $\varphi$ (an angular 
variable). Now, we ask for the operator that produces 
a cyclic permutation, within a phase shift factor, of the complete set 
of states
$\vert\varphi_1\rangle,\cdots,\vert\varphi_N\rangle$, that is, the operator
$\Delta$ that yields
$$\Delta\vert\varphi_j\rangle=\cases{\displaystyle e^{i\gamma_{j+1}}
\vert\varphi_{j+1}\rangle,&$j\ne N$,\cr\noalign{\vskip .5truecm}
\displaystyle e^{i\gamma_1}\vert\varphi_{1}\rangle,&$j=N$.\cr}$$
\noindent
Therefore, the representation of $\Delta$ in the $\varphi$--basis is the
matrix of elements
$\langle\varphi_j\vert\Delta\vert\varphi_k\rangle=\Delta_{jk}$
given by  
$$(\Delta_{jk})=\pmatrix{0&0&0&\cdots&0&e^{i\gamma_1}\cr
                      e^{i\gamma_2}&0&0&\cdots&0&0\cr
                      0&e^{i\gamma_3}&0&\cdots&0&0\cr
                      \vdots&\vdots&\vdots&\ddots&\vdots&\vdots\cr
                      0&0&0&\cdots&0&0\cr
                      0&0&0&\cdots&e^{i\gamma_N}&0\cr},\leqno(10)$$
yielding that 
$$\Delta^N=e^{i\Gamma} 1,\leqno(11)$$
where $\Gamma=\sum_{j=1}^N\gamma_j$, and 1 is the identity matrix
of dimension $N$. Since the determinant of $\Delta$ is
$$\det\Delta=(-1)^{N+1}e^{i\Gamma},$$
$\Delta$ represents a proper rotation and generates a 
finite subgroup of the rotation group if 
$$\Gamma=\cases{\displaystyle 2l\pi,&$N$ odd,\cr\noalign{\vskip .5truecm}
\displaystyle (2l+1)\pi,&$N$ even,\cr}$$
with $l$ integer. In such a case, (11) becomes
$$\Delta^N=\cases{\displaystyle 1,&N \hbox{odd,}\cr
\noalign{\vskip .5truecm}\displaystyle -1,&N \hbox{even,}\cr}
\leqno(12)$$
showing that the matrix set of powers of $\Delta$ is a representation of
the cyclic group of order $N$ or $2N$ for $N$ odd or even respectively
(being a two--valued representation of our finite group of rotations in
the latter case). Therefore, every eigenvalue of any power of $\Delta$ 
is a rational root of unity. Let us make clear this point because it is 
related with the main result of this section. By fixing the phase shifts 
$\gamma_j$, we can write (10) in more tractable forms, but due to (12) 
it is necessary to consider the cases of $N$ odd or even separately. 
Thus, we can take the $N\times N$ basic circulant permutation matrix 
$$\Delta_{jk}=\cases{\displaystyle \delta_{jN},&for\quad $j=1$,\cr
\noalign{\vskip .5truecm}
\displaystyle \delta_{j,k+1},&for\quad $j\ge 2$,\cr}$$
for (10) in the odd case, whereas, for $N$ even,
$$\Delta_{jk}=\cases{\displaystyle -\delta_{jN},&for $j=1$,\cr
\noalign{\vskip .5truecm}
\displaystyle \delta_{j,k+1},&for $j\ge 2$.\cr}$$
It is not difficult to see that the characteristic polynomial of $\Delta$
in the variable $\lambda$ is $(-\lambda)^N+1=0$ for any $N$, and that
the $N$--tuple of components
$e^{-ika_j}/\sqrt{N}$, $k=1,2,\cdots,N$, is the $j$--th normalized 
eigenvector 
of $\Delta$ yielding the eigenvalue $\lambda_j=e^{ia_j}$, where
$$a_j={{2\pi j}\over{N}},\quad j\in I_N,\leqno(13)$$
and
$$I_N=\cases{\displaystyle 0,\pm 1,\pm 2,\cdots,\pm n,& $N=2n+1$,\cr
\noalign{\vskip .5truecm}
\displaystyle \pm 1/2,\pm 3/2,\cdots,\pm (2n-1)/2,& $N=2n$.\cr}
$$\par
Being $\Delta$ unitary, it defines an hermitian matrix $A$ through
$$\Delta=e^{iA}\leqno(14)$$
whose eigenvalues are given by (13) (note that the trace of $A$ vanishes) 
and its eigenvectors, denoted by $\vert a_j\rangle$, are those of 
$\Delta$. Therefore, the unitary matrix diagonalizing simultaneously 
$\Delta$ and $A$ is the one with entries
$$\langle\varphi_k\vert a_j\rangle={1\over\sqrt{N}}e^{-ika_j},
\qquad k,j=1,2,\cdots,N.\leqno(15)$$
\noindent
In the usual quantum case $A$ becomes proportional to the angle of
rotation times $L_z$. To find a discrete analogue relation we need
to calculate the elements of $A$ in the $\varphi$--basis. This can be 
made through
$$A_{jk}=\langle\varphi_j\vert A\vert\varphi_k\rangle=
\sum_{l=1}^N\langle\varphi_j\vert a_l\rangle a_l
\langle a_l\vert\varphi_k\rangle=
{{2\pi}\over{N^2}}\sum_{l\in I_N}le^{-i(j-k)a_l}.$$
Note that the diagonal elements vanish all of them. For $N=2n+1$ we have 
$$A_{jk}=-{{4\pi}\over{N^2}}i\sum_{l=1}^n l\sin{{2\pi l(j-k)}\over{N}},$$
whereas for $N=2n$,
$$A_{jk}=-{{4\pi}\over{N^2}}i\sum_{l=1}^n(l-1/2)\sin
{{2\pi (l-1/2)(j-k)}\over{N}}.$$
These sums can be calculated easily [9] and, for $N$ odd or even, 
they become
$$A_{jk}=\langle\varphi_j\vert A\vert\varphi_k\rangle=
\cases{\displaystyle 0,& $j=k$,\cr\noalign{\vskip .5truecm}
\displaystyle {{i(-1)^{j+k}\pi/N}\over{\sin{{\pi (j-k)}\over{N}}}},
& $j\ne k$.\cr}\leqno(16)$$\par
We proceed now to establish a relation between $A$ and a $N$--dimensional 
matrix representation of the derivative of trigonometric polynomials.
We choose the odd case $N=2n+1$ first.\par
\vskip1cm
\leftline{\bf 3.1 Odd case}\par
Let $\taug_n$ be the space of 
trigonometric polynomials of degree at most $n$. It is well known 
(see for example [10]) that
any trigonometric polynomial $f\in\taug_n$ can be uniquely determined
by its values at $2n+1$ arbitrary points 
$-\pi<x_1<x_2<\cdots<x_{2n+1}\le\pi$, (we change our notation for 
simplicity) through the formula
$$f(x)=\sum_{k=1}^{2n+1}f(x_k){{t_k(x)}\over{t_k(x_k)}},$$
where the polynomials $t_k\in\taug_n$ are given by
$$t_k(x)=\prod_{l\ne k}^{2n+1}\sin{{(x-x_l)}\over 2}.\leqno(17)$$
To reach our goal we need to calculate ${{df(x)}\over{dx}}$ at the nodes.
The differentiation of (17) and algebraic manipulation of the result
yields 
$$t'_k(x_j)=\cases{\displaystyle t'(x_k)\mathop{{\sum}'}\limits_{l=1}^{2n+1}
\cot{{(x_k-x_l)}\over 2},& $j=k$,\cr
\noalign{\vskip .5truecm}
\displaystyle t'(x_j)\csc{{(x_j-x_k)}\over 2},& $j\ne k$,\cr},$$
where
$$t(x)=t_k(x)\sin{{(x-x_k)}\over2}=
\prod_{l=1}^{2n+1}\sin{{(x-x_l)}\over2}.$$
Therefore, $f'(x_j)$ takes the form
$$f'(x_j)={1\over2}t'(x_j)\mathop{{\sum}'}\limits_{k=1}^{2n+1}
{{f(x_k)}\over{\sin{{(x_j-x_k)}\over 2}t'(x_k)}}+
{1\over2}f(x_j)\mathop{{\sum}'}\limits_{k=1}^{2n+1}
\cot{{(x_j-x_k)}\over 2}.\leqno(18)$$
Note that $f'(x)$ is again an element of $\taug_n$ and that this equation 
has the structure of (4) with the definitions
$$D=D_\varphi=T\tilde{D}T^{-1},$$
$$\tilde{D}_{ij}=\cases{\displaystyle\mathop{{\sum}'}\limits_{l=1}^N
{1\over2}\cot{{(x_i-x_l)}\over 2},&{$i=j$},\cr\noalign{\vskip .5truecm}
\displaystyle {1\over2}\csc{{(x_i-x_j)}\over 2}, &{$i\not=j$},\cr}
\hskip 1.7truecm
\displaystyle T_{ij}=t'(x_i)\delta_{ij},\qquad i,j=1,N.\leqno(19)$$
Let choose the nodes such that
$$x_j=-\pi+{{2\pi j}\over N}, \quad j=1,2,\cdots,N.\leqno(20)$$
Then, it is not difficult to see that regardless the parity of $N$, 
the product (17) and the sum given by $\tilde{D}_{jj}$, evaluated at 
(20), satisfy the equations
$${{t_j(x_j)}\over{t_k(x_k)}}={{t'(x_j)}\over{t'(x_k)}}=(-1)^{j+k},$$
and
$$\tilde{D}_{jj}=\mathop{{\sum}'}\limits_{k=1}^{N}
\cot{{(j-k)\pi}\over N}=0,\quad j=1,2,\cdots,N.$$
Therefore, substituting these expressions in (18) and taking into 
account(16), we obtain for this choice of nodes and $N=2n+1$,
$$A=i{{2\pi}\over N}D_\varphi=-{{2\pi}\over N}L_z,\leqno(21)$$
where we have defined the $N$-dimensional matrix $L_z$ as
$$L_z=-iD_\varphi\leqno(22)$$
(we take $\hbar=1$). Therefore, the equation (14) for the discrete 
rotation operator $\Delta$ becomes
$$\Delta=e^{\displaystyle {-{{2\pi}\over N}D_\varphi}}=
e^{\displaystyle {-i\epsilon L_z}},\leqno(23)$$
where $\epsilon=2\pi/N$. Like it comes out in the continuum case, 
the argument of the exponential is proportional to the angle of 
rotation (in this case $\epsilon$) times the derivative $D_\varphi$. 
Thus, the interpretation of $L_z$ given by (22) as generator of 
discrete rotations follows immediately.
On account of (21), (13) and (15), $L_z$ has the $2n+1$ eigenvalues
$$\{-n,-n+1,\cdots,-1,0,1,\cdots,n-1,n\},$$
and the normalized eigenvectors $\vert m\rangle$, 
$m=-n,\cdots,n$ whose components, in the $\varphi$--basis are
$$\langle\varphi_k\vert m\rangle={1\over\sqrt{N}}e^{-im\varphi_k},$$
where $\varphi_k=k\epsilon$, $k=1,2,\cdots,N$.\par
The fact that the quantum rule for the $z$--component of the angular
momentum can be projected in a finite--dimensional space maintaining
the form it has in the continuum case (a consequence of the use of 
$D_\varphi$ as a projection of ${{d}\over{d\varphi}}$), reinforces 
the possibility of the construction of a finite--dimensional algebra 
for Quantum Mechanics as it appears in other problems [11]--[12].\par
We consider now the case $N=2n$, but before we have an important remark.
According to the uniqueness of the representation of trigonometric 
polynomials in terms of the Dirichlet kernel evaluated at differences 
of (20) (see for example [10]) we have that $D_\varphi$ equals to
$(2n+1)/2$ times the derivative of the Dirichlet kernel evaluated at 
$x_j-x_k$. On the other hand, it should be pointed out the dependence 
of the explicit form of the matrix representation of ${{d}\over{d\varphi}}$ 
on the type of projection functions; this means that it is possible 
to construct other matrices representing the derivative 
${{d}\over{d\varphi}}$, if we restrict the points to be in $(0,\pi]$ 
and project on the cosine polynomials.\par
\vskip1cm\goodbreak
\leftline{\bf 3.2. Even case}\par
Essentially, the case of $N=2n$ differs from the odd case only by the
space of functions where the projection takes place. Let us begin by
considering a function of the form
$$g(x)=\sin(x/2)f(x),\leqno(24)$$
where $f\in\taug_{n-1}$. Let $-\pi<x^*_1<x^*_2<\cdots<x^*_{2n-1}\le\pi$ be
$N-1$ arbitrary points, different from zero. We can interpolate $f$ at 
these nodes to yield
$$g(x)=\sum_{k=1}^{N-1}f(x^*_k){{\sin(x/2)t^*_k(x)}\over{t^*_k(x^*_k)}}=
\sum_{k=1}^{N-1}\big[\sin(x^*_k/2)f(x^*_k)\big]
{{\sin(x/2)t^*_k(x)}\over{\sin(x^*_k/2)t^*_k(x^*_k)}},$$
that is,
$$g(x)=\sum_{k=1}^{N-1}g(x^*_k){{\sin(x/2)t^*_k(x)}
\over{\sin(x^*_k/2)t^*_k(x^*_k)}}.\leqno(25)$$
The functions $t^*_k(x)$ are Gaussian polynomials, given by a product like 
(17) at the $N-1$ nodes $x^*_j$. Now, let us consider the set of points 
formed by zero and $x^*_j$, $j=1,2,\cdots,N-1$, and denote them by 
$x_j$, $j=1,2,\cdots,2n$. Let $t_k(x)$ be the corresponding basic 
interpolatory polynomials. Then, taking into account that $g(0)=0$, 
(25) becomes
$$g(x)=\sum_{k=1}^{2n}g(x_k){{t_k(x)}\over{t_k(x_k)}},\leqno(26)$$
and we have an interpolation formula for functions of the form (24)
at the points $x_j$. Therefore, most of the reasonings made in the
odd case can be applied to (26) to yield a formula similar to (18)
for $g(x)$, so that, for $2n$ points given by any $2n-1$ different 
points of $(-\pi,\pi]$ and zero, the matrix
$$D=D_\varphi=T\tilde{D}T^{-1},\leqno(27)$$
with the definitions (19) is an exact representation of the derivative
of functions given by (24). Moreover, taking into account that
$\cos x/2=\sin[(x+\pi)/2]$ and that $\pi$ can be taken as one of our nodes 
and that this function vanishes just at $x=\pi$, we can manipulate the 
function $h(x)=\cos(x/2)f(x)$ in the same form as we did with (24) to 
conclude that the $k$-th power of (27), constructed with a set of $2n$ 
distinct points of $(-\pi,\pi]$ that includes zero and $\pi$, is a 
$2n$--dimensional matrix projection of the $k$-th derivative on the 
subspace of functions of the type
$$e^{ix/2}f(x),$$
where $f\in\taug_{n-1}$.
Now, if we restrict these nodes to be equidistant as in (20), the same
formulas (21)-(23) for the relations between $A$, $D_\varphi$, $L_z$, 
and $\Delta$ are yielded, but now for $N=2n$. This case makes up a 
two--valued or spin representation of the finite group of rotations.\par
Using the same definitions of the preceding case, we have that
$L_z$ has the $2n$ eigenvalues
$$\{-(2n-1)/2,-(2n-3)/2,\cdots,-1/2,1/2,\cdots,(2n-3)/2,(2n-1)/2\},$$
and the normalized eigenvectors $\vert m\rangle$, 
$m=-(2n-1)/2,\cdots,(2n-1)/2$ are given again by 
$$\langle\varphi_k\vert m\rangle={1\over\sqrt{N}}e^{-im\varphi_k},$$
where $\varphi_k=k\epsilon$, $k=1,2,\cdots,N$.\par
To end this section let us remark that the validity of formulas 
(21)--(23) for both $N$ odd and even shows one discrete formulation 
of rotations in terms of an finite--dimensional matrix representation 
of the derivative of certain periodic functions that can be associated, 
according to quantum postulates, to one component of the angular momentum 
operator projected in ${\ce}^N$, giving thus, a finite subgroup of the 
rotation group that shares some of the properties of the full group.\par
\bigskip\goodbreak
\noindent
\leftline{\bf 4. The numerical eigenproblem of $L^2$.}\par
Our purpose in this section is to obtain a finite--dimensional matrix 
representation of the angular momentum eigenproblem of a system described 
by three classical degrees of freedom by applying the results of sections 
2 and 3.\par
As usual, we choose the spherical variables $\theta$ and $\varphi$ to
describe the problem. In order to apply the results of section 2,
we need two matrices: $D_\theta$ of dimension $N$, and $D_\varphi$ of
dimension $M$, to represent $d/d\theta$ and $d/d\varphi$ 
respectively. Since $-\pi<\varphi\le\pi$ and the 
functions to represent (the spherical harmonics $Y^m_n(\theta,\varphi)$)
are trigonometric polynomials, we will use as matrix $D_\varphi$, the one 
given by (19) with the change of notation $x_j\rightarrow\varphi_j$, 
where the $M$ points $\varphi_j$ are given by formula (20):
$$\varphi_j=-\pi+{{2\pi j}\over M}, \quad j=1,2,\cdots,M.$$
Concerning the variable $\theta$, we have several alternatives to choose 
a $N\times N$ matrix for $D_\theta$. Among these, we present only two
of them yielding exact results at the nodes. The first one follows the
ideas given here and the second is taken from [13].\par
\vskip1cm\goodbreak
\leftline{\bf 4.1. A matrix for $L^2$}\par
Since $0<\theta\le\pi$, we can take any set of $N$ distinct points 
$\theta_j$ of $(0,\pi)$ to construct the matrix $D_\theta$ according to 
(19). Besides, the fact that $Y^m_n(\theta,\varphi)$ are defined for 
integral indexes excludes the use of the spin representations, {\it i.e.}, 
the matrices $D_\varphi$ and $D_\theta$ should be constructed with an odd 
number of points.\par
Now, according to the results of section 2, the differential eigenvalue
problem for $f_s(\theta,\varphi)=Y^m_n(\theta,\varphi)$,
$${{\partial^2 f_s}\over{\partial^2\theta}}+\cot\theta
{{\partial f_s}\over{\partial\theta}}+{1\over{\sin^2\theta}}
{{\partial^2 f_s}\over{\partial^2\varphi}}=-\lambda_sf_s=
-n(n+1)f_s,$$
takes the matrix form
$${\lh}^2{\fh}^*_s=\lambda^*_s{\fh}^*_s,\qquad s=1,2\cdots,NM,\leqno(28)$$
where ${\lh}^2$ is  the $NM\times NM$ matrix given by
$${\lh}^2=-\big[{\dhtet}^2+\cot(\teth){\dhtet}+\sin^{-2}({\teth})
{\dhfi}^2\big],\leqno(29)$$
${\fh}^*_s\in{\ce}^{NM}$ and $\lambda^*_s$ is in general a complex number.
The matrices of (29) are given by
$${\teth}=1_M\otimes\Theta,\quad
{\dhtet}=1_M\otimes D_\theta,\quad
{\dhfi}=D_\varphi\otimes 1_N,$$
where $\Theta$ is a diagonal matrix with entries 
$\Theta_{jk}=\theta_j\delta_{jk},$ and $1_K$ is the identity matrix
of dimension $K$. Thus, (29) can be rewritten as
$$-{\lh}^2=1_M\otimes[D_\theta^2+\cot(\Theta)D_\theta]+
D_\varphi^2\otimes\sin^{-2}(\Theta).\leqno(30)$$\par
Since $Y^m_n(\theta,\varphi)$ is the (tensor) product of the associated 
Legendre functions $P^m_n(\theta)$ and $e^{im\theta}$ and, on the other
hand, we know (from the results of section 3) that the 
degree of the polynomial that can be differentiated through (19) 
yielding exact values at $N$ nodes ($N$ odd) is at most $(N-1)/2$, 
formula (28) reproduces exactly the first
$$\sum_{l=0}^{(N-1)/2}(2l+1)=\Big({{N+1}\over2}\Big)^2$$
eigenvalues and the corresponding unnormalized functions 
$Y^m_n(\theta,\varphi)$ evaluated at the nodes $(\theta_j,\varphi_k)$, 
provided $M\ge N$. In this case ${\lh}^2$ will have necessarily 
$(N+1)^2/4$ real eigenvalues given by 
$$\lambda^*_s=n(n+1),$$
ordered according to 
$$s=n^2+n+m+1,\quad n=0,1,\cdots,(N-1)/2,\quad m=-n,-n+1,\cdots,n-1,n.$$
The eigenvector ${\fh}^*_s$ corresponding to $\lambda^*_s$, has the 
components $f_{rs}^*$ given by 
$$f_{rs}^*=c_{nm}P^m_n(\theta_j)e^{imk\epsilon},$$
where $\epsilon=2\pi/M$, $c_{nm}$ is a normalization constant, 
$\theta_j\in(0,\pi)$, and the relation between $r$, $j$ and $k$, is 
given by (6).\par
It is possible to choose the $\theta$--nodes in such a way that
${\lh}^2$ becomes a positive semidefinite matrix (save to a similarity
transformation). To see this, note that $-D_\varphi^2$ is positive 
semidefinite whereas $\sin^{-2}(\Theta)$ is positive definite since 
$\theta_j\in(0,\pi)$, $j=1,2,\cdots,N$.
Therefore, according to (30) we only have to find the conditions
which make the matrix
$$T_\theta^{-1}L^2_\theta T_\theta=-\tilde{D}_\theta^2-\cot(\Theta)
\tilde{D}_\theta\leqno(31)$$
positive semidefinite. To this end, let us separate the main diagonal 
of $D_\theta$ by writing 
$$D_\theta=T_\theta D^*T_\theta^{-1}+d,$$
where
$$D^*_{ij}=\cases{0,&{$i=j$},\cr\noalign{\vskip .5truecm}
\displaystyle {1\over2}\csc{{(\theta_i-\theta_j)}\over 2}, &{$i\not=j$},\cr}
\hskip 1.7truecm
\displaystyle d_{ij}=\delta_{ij}\displaystyle\mathop{{\sum}'}
\limits_{l=1}^N{1\over2}\cot{{(\theta_i-\theta_j)}\over 2}$$
\par\noindent
[{\it cf.} (19)], and $T_\theta$ is given by (19). Thus,
$$L^2_\theta=-T_\theta [{D^*}^2+D^*d+dD^*+d^2+\cot(\Theta)D^*+\cot(\Theta)d]
T_\theta^{-1}.\leqno(32)$$
If we can find points $\theta_j$ such that $d=-\cot(\Theta)/2$, {\it i.e.},
$$\mathop{{\sum}'}\limits_{l=1}^N\cot{{(\theta_j-\theta_l)}\over 2}=
-\cot(\theta_j)\leqno(33)$$
[{\it cf.} (3)], equation (32) becomes
$$L^2_\theta=T_\theta (-{D^*}^2+dD^*-D^*d+d^2)T_\theta^{-1}=
T_\theta(-D^*+d)(D^*+d)
T_\theta^{-1}=T_\theta \tilde{D^t_\theta} \tilde{D_\theta} T_\theta^{-1},$$
where $\tilde{D^t_\theta}$ is the transpose of $\tilde{D_\theta}$.
It remains to show
that a solution of (33) always exists. Since $\cot x$ is the 
logarithmic derivative of $\sin x$, Eq. (33) is the condition 
for a critical point of the function of $N$ variables 
$$U(z)=U(z_1,z_2,\cdots,z_N)=\prod_{k=1}^N\sin(z_k)
\prod_{i>j}^N\sin{{(z_i-z_j)}\over 2},$$
where $0\le z_j\le\pi$, $j=1,2,\cdots,N$.
The existence (and uniqueness) of the solution can be proved along
the same lines given in [14]. 
It is worth to be noticed that the similarity transformation is not 
essential for an eigenproblem like (28) since the
similarity matrices corresponding to $\theta$ and $\varphi$ can be 
collected into a $NM\times NM$ diagonal matrix ${\sh}$ to write (28) 
in the form
$${\lh}_P^2{\gh}_s=\lambda^*_s{\gh}_s,$$
where ${\gh}_s={\sh}^{-1}{\fh}^*_s$ and ${\lh}_P^2$ is positive 
semidefinite. Thus, if we construct $L^2_\theta$ with the set of nodes 
satisfying (33), ${\lh}^2$ is a positive semidefinite matrix (save a 
similarity transformation), a necessary property from the numerical and 
physical point of view in any projection scheme.\par
\vskip1cm\goodbreak
\leftline{\bf 4.2. Other matrix for $L^2$}\par
In [13], a matrix representation of the derivative for trigonometric 
polynomials of definite parity is given and, therefore, it can also be 
used to construct a representative of (31), yielding a matrix for $L^2$ 
with a higher degree of approximation than (29). The cost we have to 
afford for this is the lacking of simplicity: we can not use a single 
matrix for ${{d}\over{d\theta}}$ to be substituted directly in (31)
unless we
accept to loose precision in the results (see [13]). According to 
this scheme, the representation of a differential operator 
is formed following a given rule where certain matrix {\dcal}
is involved. Such a matrix is constructed with $N$ arbitrary distinct
points $\theta_j\in (0,\pi)$ through a formula similar to (19)
$${\dcal}={\dcal}_\theta=S\tilde{\dcal}S^{-1},$$
$$\tilde{\dcal}_{ij}=\cases{\displaystyle\mathop{{\sum}'}\limits_{l=1}^N
\cot(\theta_i-\theta_l),&{$i=j$},\cr\noalign{\vskip .5truecm}
\displaystyle \cot(\theta_i-\theta_j), &{$i\not=j$},\cr}
\hskip 1.7truecm \displaystyle S_{ij}=\delta_{ij}\prod_{l\ne j}^N
\sin(\theta_j-\theta_l).$$\par
The form that $L^2_\theta$ adopts in this case is
$$L^2_\theta=-{\dcal}_\theta^2-\cot(\Theta){\dcal}_\theta+
NSOS^{-1},\leqno(34)$$
where $\Theta$ is again a diagonal matrix with entries 
$\Theta_{jk}=\theta_j\delta_{jk}$ and $O$ is a projection matrix with
ones everywhere, {\it i.e.}, $O_{jk}=1$. According to [13], the degree of 
approximation of (34) is higher than that of (31). While the former 
yields exact results at $N$ nodes ($N$ odd or even) for trigonometric 
polynomials (of definite parity) of degree $N$, the latter produces 
exact results for polynomials of degree $(N-1)/2$ ($N$ odd). Thus,
the use of (34) in (30) gives the matrix 
$$-{\lh}^2=1_M\otimes[{\dcal}_\theta^2+\cot(\Theta){\dcal}_\theta-
NSOS^{-1}]+D_\varphi^2\otimes\sin^{-2}(\Theta)\leqno(35)$$
with a higher degree of approximation: if $N$ is an odd integer and 
$M=2N+1$, then (35) produces the first
$$\sum_{l=0}^N(2l+1)=(N+1)^2$$
exact eigenvalues (and eigenvectors) while (30) yields only the first 
$(N+1)^2/4$ for the same values of $N$ and $M$.\par
\vskip1cm\goodbreak
\leftline{\bf 5. Final remarks}\par
Summarizing, we have found a finite--dimensional representation of the 
square of quantum angular momentum with the following properties:\par
\item{1. }The coordinate representation of the $z$--component of the 
angular momentum operator is maintained in this discrete scheme and the
projection of $L_z$ is the generator of a finite subgroup of the group of
rotations.
\item{2. }The projection of $L^2$ is given in terms of finite--dimensional 
representations of the partial derivatives according to the well-known 
formula for $L^2$.
\item{3. }The spectrum of this projection contains the first eigenvalues 
of $L^2$ (counting the degeneracy) in such a form that the whole spectrum 
of $L^2$ can be reobtained when the number of nodes tends to infinity. 
\item{4. }The eigenvectors of this projection, corresponding to the exact 
eigenvalues of $L^2$, can be converted into the exact eigenfunctions through 
an interpolation at the nodes. Again, this process yields the complete set 
of eigenfunctions of $L^2$ when the dimension of ${\lh}^2$ tends to infinity.
\par
Finally, we note that these properties makes ${\lh}^2$ suitable for
numerical applications to quantum problems as it will be shown in a
subsequent work.\par
\vskip .6truecm
\leftline{\bf Acknowledgments}
RGC wants to thank Prof. O. Obreg\'on and Prof. L. Sabinin
for very useful discussions. This research has been partially supported
by Consejo Nacional de Ciencia y Tecnolog\'{\i}a (CONACYT). 

\vfill\eject
\baselineskip 12pt
\vskip 2truecm
\centerline{\bf REFERENCES}
\bigskip
\item{[1] }F. Calogero, Lett. Nuovo Cimento {\bf 35}, 273 (1982).
\item{[2] }F. Calogero and E. Franco, Nuovo Cimento B {\bf 89}, 161 (1985).
\item{[3] }M. Bruschi, R.G. Campos and E. Pace, Nuovo Cimento B {\bf 105}, 
131 (1990).
\item{[4] }R.G. Campos, Bol. Soc. Mat. Mexicana, {\bf 3}, 279 (1997)
\item{[5] }R.G. Campos y R. Mu\~noz B., Rev. Mex. F\'{\i}s. {\bf 36},
1 (1990).
\item{[6] }F. Calogero, J. Math. Phys. {\bf 34}, 4704 (1993)
\item{[7] }E.W. Cheney, {\it Multivariate approximation theory: selected
topics} (SIAM Publications, Philadelphia, Pennsylvania, 1986)
\item{[8] }F. Calogero, Lett. Nuovo Cimento {\bf 38}, 453 (1983).
\item{[9] }I.S. Gradshteyn and I.M. Ryzhik, {\it Table of integrals, series,
and products} (Academic Press, London, 1994),  5th. Ed.
\item{[10] }A.F. Timan, {\it Theory of Approximation of functions of a real
variable} (Dover Publications, Inc., New York, 1994)
\item{[11] }R.G. Campos, Rev. Mex. Fis, {\bf 29}, 217 (1983).
\item{[12] }R.G. Campos, Rev. Mex. Fis., {\bf 32}, 379 (1986).
\item{[13] }F. Calogero, Lett. Nuovo Cimento {\bf 39}, 305 (1984).
\item{[14] } T. Popoviciu, Bull. Math. Soc. Roumaine Sci., {\bf 38},
73 (1936).

\end